\documentclass[10pt]{article}
\usepackage{typearea}\typearea{14}
\usepackage{epsfig,amsmath,amsfonts,amssymb,upgreek,dsfont}
\hypersetup{hidelinks}

\makeatletter
\@addtoreset{equation}{section}
\makeatother

\makeatletter
\long\def\@makecaption#1#2{{\small
\advance\leftskip1cm
\advance\rightskip1cm
\vskip\abovecaptionskip
\sbox\@tempboxa{#1: #2}%
\ifdim \wd\@tempboxa >\hsize
 #1: #2\par
\else
\global \@minipagefalse
\hb@xt@\hsize{\hfil\box\@tempboxa\hfil}%
\fi
\vskip\belowcaptionskip}}
\makeatother
\def\eq#1\en{\begin{equation}#1\end{equation}}  
\def\eqa#1\ena{\begin{align}#1\end{align}}
\def\eqg#1\eng{\begin{gather}#1\end{gather}}
\newcommand{\lb}[1]{\label{e:#1}}
\newcommand{\rlb}[1]{\eqref{e:#1}} 
\newcommand{\nl}{\notag\\}

\newcommand{\bkt}[1]{\left\langle#1\right\rangle}
\newcommand{\sbkt}[1]{\langle#1\rangle}

\newcommand{\sumtwo}[2]%
{\mathop{\sum_{#1}}_{#2}}
\newcommand{\sumthree}[3]%
{\mathop{\mathop{\sum_{#1}}_{#2}}_{#3}}
\newcommand{\sumfour}[4]%
{\mathop{\mathop{\mathop{\sum_{#1}}_{#2}}_{#3}}_{#4}} 
\newcommand{\prodtwo}[2]%
{\mathop{\prod_{#1}}_{#2}}
\newcommand{\mintwo}[2]%
{\mathop{\min_{#1}}_{#2}}
\newcommand{\maxtwo}[2]%
{\mathop{\max_{#1}}_{#2}}
\newcommand{\maxthree}[3]%
{\mathop{\mathop{\max_{#1}}_{#2}}_{#3}}
\newcommand{\limtwo}[2]%
{\mathop{\lim_{#1}}_{#2}}
\newcommand{\suptwo}[2]%
{\mathop{\sup_{#1}}_{#2}}
\newcommand{\supthree}[3]%
{\mathop{\mathop{\sup_{#1}}_{#2}}_{#3}}
\newcommand{\supfour}[4]%
{\mathop{\mathop{\mathop{\sup_{#1}}_{#2}}_{#3}}_{#4}} 
\newcommand{\inftwo}[2]%
{\mathop{\inf_{#1}}_{#2}}
\newcommand{\infthree}[3]%
{\mathop{\mathop{\inf_{#1}}_{#2}}_{#3}}
\newcommand{\inffour}[4]%
{\mathop{\mathop{\mathop{\inf_{#1}}_{#2}}_{#3}}_{#4}} 

\newcommand\calP{{\cal P}}

\newcommand{\bsp}{\boldsymbol{p}}

\newcommand{\bsr}{\boldsymbol{r}}
\newcommand{\bss}{\boldsymbol{s}}

\newcommand{\bsx}{\boldsymbol{x}}

\newcommand{\bbR}{\mathbb{R}}

\newcommand{\Di}{\mathit{\Delta}}

\newcommand{\La}{\Lambda}
\newcommand{\LaL}{\Lambda_{\rm L}}
\newcommand{\LaR}{\Lambda_{\rm R}}
\newcommand{\bsR}{\boldsymbol{R}}
\newcommand{\bsP}{\boldsymbol{P}}
\newcommand{\NL}{N_{\rm L}}
\newcommand{\NR}{N_{\rm R}}
\newcommand{\CLR}{\mathcal{C}_{\rm LR}}
\newcommand{\ZLR}{\tilde{Z}_{\rm LR}}
\newcommand{\Htrap}{H^{(\lambda)}_{\rm trap}}
\newcommand{\utrap}{u_{\rm trap}}
\newcommand{\vrep}{v_{\rm rep}}

\begin{document}

\begin{flushright}
\footnotesize
Not for publication in a journal.
\end{flushright}

\noindent
{\LARGE\bf 
The best answer to the puzzle of Gibbs about $N!$!}

\vspace{1.5mm}

\noindent{\Large\bf
A note on the paper by Sasa, Hiura, Nakagawa, and Yoshida}

\renewcommand{\thefootnote}{\fnsymbol{footnote}}
\medskip\noindent
Hal Tasaki\footnote{%
Department of Physics, Gakushuin University, Mejiro, Toshima-ku, 
Tokyo 171-8588, Japan.
}
\renewcommand{\thefootnote}{\arabic{footnote}}
\setcounter{footnote}{0}

\begin{quotation}
\small\noindent
In a recent paper, \cite{SHNY}, Sasa, Hiura, Nakagawa, and Yoshida showed that a natural extension of the minimum work principle to small systems uniquely determines the factor $N!$ that arrises in relations connecting statistical mechanical functions (such as the partition function) and thermodynamic functions (such as the free energy).
We believe that this provides us with the clearest answer to the ``puzzle'' in classical statistical mechanics that goes back to Gibbs.

Here we attempt at explaining the theory of Sasa, Hiura, Nakagawa, and Yoshida \cite{SHNY} by using a process discussed by Horowitz and Parrondo \cite{HP} in a different context.
Although the content of the present note should be obvious to anybody familiar with both \cite{SHNY} and \cite{HP}, we believe it is useful to have a commentary that presents the same theory from a slightly different perspective.

The present note is written in a self-contained manner.
We only assume basic knowledge of classical statistical mechanics and thermodynamics.
We nevertheless invite the reader to refer to the original paper \cite{SHNY} for background, references, and related discussions, as well as the original thoughts.

\medskip\noindent
There is a 20 minutes lecture video in which I discuss the essence of the present note.\\
\url{https://youtu.be/gG17ttXeDhA}\\
I believe it is a good idea to watch the video before start reading this.
\end{quotation}

\tableofcontents

\section{Introduction and preliminary considerations}
\subsection{Setting and the motivation}
We consider a classical system of $N$ identical point-like particles confined in a box $\La$.
We do not assume that $N$ is large since we wish to treat small statistical mechanical systems as well as macroscopic ones.
We label the particles by indices $j=1,\ldots,N$ and denote their positions and momenta as $\bsr_1,\ldots,\bsr_N\in\La$ and $\bsp_1,\ldots,\bsp_N\in\bbR^3$, respectively.
We express the coordinates collectively as $\bsR=(\bsr_1,\ldots,\bsr_N)\in\La^N$ and $\bsP=(\bsp_1,\ldots,\bsp_N)\in\bbR^{3N}$, and also write $d\bsR=d^3\bsr_1\cdots d^3\bsr_N$ and $d\bsP=d^3\bsp_1\cdots d^3\bsp_N$.
Note that the identity of particles guarantees that one can freely redefine the particle labels.

Consider, for simplicity, a typical Hamiltonian\footnote{%
The whole discussion can readily be extended to a much more general class of Hamiltonians.}
\eq
H(\bsR,\bsP)=\sum_{j=1}^N\frac{|\bsp_j|^2}{2m}+\sum_{j=1}^Nu(\bsr_j)+\frac{1}{2}\mathop{\sum_{j,k=1}^N}_{(j\ne k)}v(\bsr_j,\bsr_k),
\lb{H}
\en
with a single-particle potential $u(\bsr)$ and two-particle interaction $v(\bsr,\bsr')$.
The equilibrium state of the system is described by the canonical distribution
\eq
\calP_{\beta,H}(\bsR,\bsP)=\frac{e^{-\beta H(\bsR,\bsP)}}{Z(\beta,H)},
\lb{r}
\en
for $(\bsR,\bsP)\in\La^N\times\bbR^{3N}$, where the inverse temperature $\beta>0$ is fixed throughout the present note.
The partition function $Z(\beta,H)$ is the normalization factor defined by
\eq
Z(\beta,H)=\int d\bsR\,d\bsP\,e^{-\beta H(\bsR,\bsP)}.
\lb{Z}
\en
For an arbitrary observable $O(\bsR,\bsP)$, its equilibrium expectation value is given by
\eq
\sbkt{O}_{\beta,H}=\int d\bsR\,d\bsP\,O(\bsR,\bsP)\,\calP_{\beta,H}(\bsR,\bsP).
\lb{O}
\en

It is also standard to define the Helmholtz free energy corresponding to the equilibrium state \rlb{r} by
\eq
F(\beta,H)=-\frac{1}{\beta}\log \frac{1}{c^N\,N!}\,Z(\beta,H),
\lb{F}
\en
where the positive constant $c$ is arbitrary but usually chosen as $c=h^3$ with the Planck constant $h$.
One may accept the expression \rlb{F} as a given definition, but we find it desirable to have a justification based on some physical reasonings.\footnote{ 
A formal derivation of \rlb{F} may be provided by quantum statistical mechanics.
Take a quantum system of $N$ identical bosons or fermions with the Hamiltonian
\eq
\hat{H}=\sum_{j=1}^N\frac{|\hat{\bsp}_j|^2}{2m}+\sum_{j=1}^Nu(\hat{\bsr}_j)+\frac{1}{2}\mathop{\sum_{j,k=1}^N}_{(j\ne k)}v(\hat{\bsr}_j,\hat{\bsr}_k),
\en
which is the ``quantization'' of above \rlb{H}.
Then it can be shown that the quantum mechanical partition function is approximated in the semiclassical limit as
\eq
Z_{\rm qm}(\beta,\hat{H})=
\operatorname{Tr} e^{-\beta\hat{H}}\simeq \frac{1}{h^{3N}\,N!}\,Z(\beta,H),
\en
which is nothing but the argument of the logarithm in \rlb{F}.

We thus see that the desired formula \rlb{F} is derived if one assumes that $F=-\beta^{-1}\log Z_{\rm qm}$.
But note that there are no physical reasons to justify this last assumption about $F$.
See also section~\ref{s:standard}.
} 
We in particular look for an argument that makes use of the operational point of view common in thermodynamics.

\subsection{General properties of $Z(\beta,H)$}
Before proceeding, we summarize two important properties of the partition function $Z(\beta,H)$.

First, we recall that the energy expectation value in the equilibrium state is compactly expressed in term of the partition function as
\eq
\sbkt{H}_{\beta,H}=-\frac{\partial}{\partial\beta}\log Z(\beta,H).
\lb{HZ}
\en
This is easily verified by combining the definitions \rlb{r}, \rlb{Z}, and \rlb{O}.

Secondly, we note that the work in a quasi-static isothermal process is written in terms of the partition function.
To be precise let $H_\alpha(\bsR,\bsP)$ be an arbitrary smooth one-parameter family of Hamiltonians.
Suppose that the system starts from the equilibrium state of $H_0(\bsR,\bsP)$ and the parameter $\alpha$ is varied indefinitely slowly from 0 to 1 while the system is in touch with a heat bath at inverse temperature $\beta$.
We imagine that the Hamiltonian is modified by an external agent who causes this thermodynamic process.
Then the total work that the agent does to the system is given by
\eq
W(\beta;H_0\to H_1)=\frac{1}{\beta}\log Z(\beta,H_0)-\frac{1}{\beta}\log Z(\beta,H_1).
\lb{WZZ}
\en
Although this is a well-known result, we derive it below for completeness.

\medskip

\noindent{\em Derivation of \rlb{WZZ}:}\/
The work required (for an external agent that modifies the Hamiltonian) to vary the parameter from $\alpha$ to $\alpha+\Di\alpha$ is given by\footnote{
It is essential here that the expectation value is with respect to the initial Hamiltonian $H_\alpha$, and hence $\Di W$ is different from the change in the energy $\Di U=\sbkt{H_{\alpha+\Di\alpha}-H_{\alpha}}_{\beta,H_{\alpha+\Di\alpha}}-\sbkt{H_{\alpha}}_{\beta,H_\alpha}$.
The difference $\Di U-\Di W$ is the heat absorbed by the system.
}
\eqa
\Di W&=\sbkt{H_{\alpha+\Di\alpha}-H_{\alpha}}_{\beta,H_\alpha}+O(\Di\alpha)^2)
\nl&=\Di\alpha\bkt{\frac{\partial H_\alpha}{\partial\alpha}}_{\beta,H_\alpha}+O(\Di\alpha)^2)
\nl&=-\Di\alpha\,\frac{1}{\beta}\frac{\partial}{\partial\alpha}\log Z(\beta,H_\alpha).
\ena
Thus the total work needed in an isothermal process in which $\alpha$ is varied indefinitely slowly from 0 to 1 is given by
\eqa
W(\beta;H_0\to H_1)&=-\int_0^1d\alpha\,\frac{1}{\beta}\frac{\partial}{\partial\alpha}\log Z(\beta,H_\alpha)
\nl&=\frac{1}{\beta}\log Z(\beta,H_0)-\frac{1}{\beta}\log Z(\beta,H_1).
\ena

\subsection{The standard premises for the free energy}\label{s:standard}
Let us discuss the standard premises for the Helmholtz free energy and its consequences.
These premises are motivated by thermodynamics.

\paragraph{The first premise}
First, the thermodynamic Helmholtz free energy satisfies the Gibbs-Helmholtz relation
\eq
U(\beta,H)=\frac{\partial}{\partial\beta}\{\beta\,F(\beta,H)\},
\lb{req1}
\en
for any $\beta$ and $H$,
where the energy $U(\beta,H)$ in the equilibrium is identified with the expectation value $\sbkt{H}_{\beta,H}$.
Comparison with the statistical mechanical relation \rlb{HZ} shows that one should set
\eq
F(\beta,H)=-\frac{1}{\beta}\log\bigl[\Phi\,Z(\beta,H)\bigr],
\lb{FZ1}
\en
where the quantity $\Phi$ does not depend on $\beta$.

\paragraph{The second premise}
Secondly, we require the validity of the minimum work principle, i.e., the work done by an external agent in a quasi-static isothermal process in which the Hamiltonian is changed from $H_0$ to $H_1$ should be given by
\eq
W(\beta;H_0\to H_1)=F(\beta;H_1)-F(\beta;H_0).
\lb{req2}
\en
From \rlb{WZZ}, we see that this is satisfied provided that $\Phi$ in \rlb{FZ1} is independent also of the Hamiltonian $H$.

Since the particle number $N$ is conserved in an isothermal process, the above considerations indicate that $\Phi$ can be a function of $N$.
We thus find that the expression of the free energy in terms of the partition function determined from the above two premises is
\eq
F(\beta,H)=-\frac{1}{\beta}\log\bigl[\Phi(N)\,Z(\beta,H)\bigr],
\lb{FZ2}
\en
where $\Phi(N)$ is a certain undetermined function of $N$.

It should be noted that the choice of the function $\Phi(N)$ does not affect physically observable quantities, i.e. the energy, the generalized forces, and the works in quasi-static processes, that can be obtained from the Helmholtz free energy.
One may say that the choice is only a matter of convention and convenience, but it is sometimes essential to use a good convention.

It is also worth noting that we have not assumed that the system is macroscopic in the above discussion.
The conclusion \rlb{FZ2} applies to small systems as well (provided that one makes these requirements).

\paragraph{The third premise}
In the third premise, on the other hand, one assumes that the system is macroscopic.
(This is hence not useful for us.)

In thermodynamics, the Helmholtz free energy expressed as a function of the temperature $T$, the volume $V$, and the amount of substance $N$ must be extensive in the sense that
\eq
F[T;\lambda V,\lambda N]=\lambda\,F[T;V,N],
\lb{ext}
\en
for any scaling factor $\lambda>0$.
To see the implication of this premise, consider the ideal gas whose Hamiltonian is \rlb{H} with $u(\bsr)=0$ and $v(\bsr,\bsr')=0$.
From an explicit computation we see that
\eq
Z(\beta,H_{\rm ideal})=V^N\biggl(\frac{2\pi m}{\beta}\biggr)^{3N/2}.
\en
Then one finds that the free energy given by \rlb{FZ2} satisfies extensivity \rlb{ext} only when one chooses
\eq
\Phi(N)=(c'\,N)^{-N},
\lb{PN1}
\en
with a constant $c'>0$.
In this case, the free energy is
\eq
F(\beta,H_{\rm ideal})=-\frac{N}{\beta}\Bigl(\log\frac{V}{N}+\frac{3}{2}\log\frac{2\pi m}{\beta}-\log c\Bigr).
\en
Because of Stirling's formula $N!\sim(N/e)^N$, one finds that the choice \rlb{PN1} is essentially the same as
\eq
\Phi(N)=\frac{1}{(c'\,e)^N\,N!},
\lb{PN2}
\en
which leads to the desired \rlb{F} with $c=c'e$, 
provided that $N\gg1$.

Note that the above discussion is meaningless for small systems, in which we are interested.
First of all, the extensivity assumption \rlb{ext} can be valid only for macroscopic thermodynamic systems.
In small systems, there always exist boundary effects and the scaling as in \rlb{ext} can be satisfied only approximately.
Secondly, we are allowed to identify $N!$ with $(N/e)^N$ only when $N$ is extremely large.

In order to justify the expression \rlb{F} also for small $N$, one needs to find a general premise that works for small systems and leads directly to the $N!$ factor, rather than $N^N$.
As we shall see in the next section, Sasa, Hiura, Nakagawa, and Yoshida came up with a simple premise that naturally leads to the desired factor $N!$ in small systems.

\section{The premise of Sasa, Hiura, Nakagawa, and Yoshida and its conseuence}
We shall introduce a quasi-static isothermal process proposed in \cite{SHNY}, which we call the  Sasa-Hiura-Nakagawa-Yoshida  (SHNY) process.
We then discuss the new premise of Sasa, Hiura, Nakagawa, and Yoshida that refines the extensivity assumption.

\subsection{The initial sate}
The initial state of the SHNY process is the equilibrium state
\eq
\calP_0(\bsR,\bsP)=\frac{e^{-\beta H_0(\bsR,\bsP)}}{Z_0(\beta)},
\lb{r0}
\en
for the standard Hamiltonian
\eq
H_0(\bsR,\bsP)=\sum_{j=1}^N\frac{|\bsp_j|^2}{2m}+\sum_{j=1}^Nu_0(\bsr_j)+\frac{1}{2}\mathop{\sum_{j,k=1}^N}_{(j\ne k)}v_0(\bsr_j,\bsr_k),
\lb{H0}
\en
with arbitrary single-particle potential $u_0(\bsr)$ and arbitrary two-particle interaction $v_0(\bsr,\bsr')$.
Of course $Z_0(\beta)$ is the partition function
\eq
Z_0(\beta)=\int d\bsR\,d\bsP\,e^{-\beta H_0(\bsR,\bsP)}.
\lb{Z0}
\en

\subsection{The final sate}
To define the final state of the SHNY process, we assume that the region $\La$ is divided into two regions, $\LaR$ and $\LaL$, by an infinitesimally thin planar wall that perfectly reflects particles.
We let $u_1(\bsr)$ with $\bsr\in\La$ be an arbitrary single-particle potential, and $v_1(\bsr,\bsr')$ with $\bsr,\bsr'\in\La$ be an arbitrary two-particle interaction with the property that $v_1(\bsr,\bsr')\ne0$ only when $\bsr,\bsr'\in\LaL$ or  $\bsr,\bsr'\in\LaR$.
In other words, particles on different subregions do not interact.

We also fix positive integers $\NL$ and $\NR$ such that $\NL+\NR=N$ and assume that there are $\NL$ particles in the region $\LaL$ and $\NR$ particles in $\LaR$.
Since particle labels are arbitrary, we may assume that particles $1,\ldots,\NL$ are in $\LaL$, and particles $\LaL+1,\ldots,N$ are in $\LaR$.
Correspondingly we write $\bsR_{\rm L}=(\bsr_1,\ldots,\bsr_{\NL})\in\LaL^{\NL}$, $\bsP_{\rm L}=(\bsp_1,\ldots,\bsp_{\NL})\in\bsR^{3\NL}$, $\bsR_{\rm R}=(\bsr_{\NL+1},\ldots,\bsr_N)\in\LaR^{\NR}$, and $\bsP_{\rm R}=(\bsp_{\NL+1},\ldots,\bsp_{N})\in\bsR^{3\NR}$.

We define the Hamiltonians as
\eqg
H_{\rm L}(\bsR_{\rm L},\bsP_{\rm L})=\sum_{j=1}^{\NL}\frac{|\bsp_j|^2}{2m}+\sum_{j=1}^{\NL}u_1(\bsr_j)+\frac{1}{2}\mathop{\sum_{j,k=1}^{\NL}}_{(j\ne k)}v_1(\bsr_j,\bsr_k),
\lb{HL}\\
H_{\rm R}(\bsR_{\rm R},\bsP_{\rm R})=\sum_{j=\NL+1}^{N}\frac{|\bsp_j|^2}{2m}+\sum_{j=\NL+1}^{N}u_1(\bsr_j)+\frac{1}{2}\mathop{\sum_{j,k=\NL+1}^{N}}_{(j\ne k)}v_1(\bsr_j,\bsr_k),
\eng
and the corresponding equilibrium states as
\eq
\calP_{\rm L}(\bsR_{\rm L},\bsP_{\rm L})=\frac{e^{-\beta H_{\rm L}(\bsR_{\rm L},\bsP_{\rm L})}}{Z_{\rm L}(\beta)},
\quad
\calP_{\rm R}(\bsR_{\rm R},\bsP_{\rm R}) =\frac{e^{-\beta H_{\rm R}(\bsR_{\rm R},\bsP_{\rm R})}}{Z_{\rm R}(\beta)},
\lb{rLrR}
\en
where the partition functions are
\eq
Z_{\rm L}(\beta)=\int d\bsR_{\rm L}\,d\bsP_{\rm L}\,e^{-\beta H_{\rm L}(\bsR_{\rm L},\bsP_{\rm L})},\quad
Z_{\rm R}(\beta)=\int d\bsR_{\rm R}\,d\bsP_{\rm R}\,e^{-\beta H_{\rm R}(\bsR_{\rm R},\bsP_{\rm R})},
\lb{ZLZR}
\en
with $d\bsR_{\rm L}=d^3\bsr_1\cdots d^3\bsr_{\NL}$, $d\bsP_{\rm L}=d^3\bsp_1\cdots d^3\bsp_{\NL}$, $d\bsR_{\rm R}=d^3\bsr_{\NL+1}\cdots d^3\bsr_N$, and $d\bsP_{\rm R}=d^3\bsp_{\NL+1}\cdots d^3\bsp_N$.
The final state of the SHNY process is the product
\eq
\calP_{\rm fin}(\bsR,\bsP)=
\calP_{\rm L}(\bsR_{\rm L},\bsP_{\rm L})\,\calP_{\rm R}(\bsR_{\rm R},\bsP_{\rm R}).
\lb{rLR}
\en

We stress that the final state $\calP_{\rm fin}$, although being a simple product of two equilibrium states, is not an equilibrium state in a strict sense.
In general, the left and the right regions may have different pressures, in which case the whole state is manifestly nonequilibrium.
More importantly, even when the pressures are equal, the state $\calP_{\rm fin}$ is distinct from the genuine equilibrium state attained in a situation where the particles can move between the two regions.
In the latter case, the particle numbers in each region exhibit fluctuation, while the particle numbers are fixed in $\calP_{\rm fin}$.
Note that the presence or the lack of particle number fluctuation is of essential physical importance especially when the particle numbers are not macroscopic.

\subsection{The SHNY process and the basic premise}
Sasa, Hiura, Nakagawa, and Yoshida \cite{SHNY} proposed to study quasi-static isothermal processes that connect the states \rlb{r0} and \rlb{rLR}.
We shall call such a process a SHNY process.

The premise of Sasa, Hiura, Nakagawa, and Yoshida, which replaces the extensivity assumption, is that the minimum work principle should also apply to SHNY processes.
To be precise, let $W_{\rm SHNY}$ denote the work required for a SHNY process.
Then one requires that
\eq
W_{\rm SHNY}=\{F(\beta,\NL,H_{\rm L})+F(\beta,\NR,H_{\rm R})\}-F(\beta,N,H),
\lb{SHNY}
\en
where $F(\beta,N,H)$, $F(\beta,\NL,H_{\rm L})$, and $F(\beta,\NR,H_{\rm R})$ are the Helmholtz free energies associated with the equilibrium states $\calP_0$, $\calP_{\rm L}$, and $\calP_{\rm R}$, respectively, defined in \rlb{r0} and \rlb{rLrR}.

In the next section, we shall see that the above work is expressed in terms of the partition functions defined in \rlb{Z0} and \rlb{ZLZR} as
\eqa
W_{\rm SHNY}=\frac{1}{\beta}\log\frac{Z_0(\beta)}{N!}-\Bigl\{\frac{1}{\beta}\log\frac{Z_{\rm L}(\beta)}{\NL!}
+\frac{1}{\beta}\log\frac{Z_{\rm R}(\beta)}{\NR!}\Bigr\}.
\lb{Wtot}
\ena
By substituting the expression \rlb{FZ2} of the free energy into the basic premise \rlb{SHNY} and comparing the result with \rlb{Wtot}, one finds that the function $\Phi(N)$ must be taken as
\eq
\Phi(N)=\frac{1}{c^N\,N!},
\lb{PNfinal}
\en
with an undetermined constant $c>0$.
We have thus recovered the standard expression \rlb{F} for the Helmholtz free energy even for systems with small $N$.

In thermodynamics, the premise \rlb{SHNY} is a well-known useful relation that follows from the minimum work principle and the additivity of the free energy.
What is essential in the proposal of Sasa, Hiura, Nakagawa, and Yoshida is that the formally identical relation is required for small systems, where the final state $\calP_{\rm fin}$ is not a strict equilibrium state.
We also recall that neither the additivity nor the extensivity is valid in small equilibrium systems.

\begin{figure}
\centerline{\epsfig{file=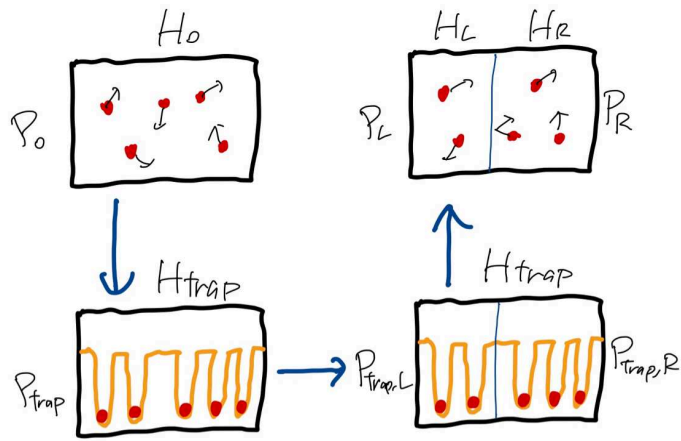,width=9truecm}}
\caption[dummy]{
A realization of the SHNY process.
It consists of the trapping process, the insertion of the wall, and the untrapping process.
Essentially the same process was discussed by Horowitz and Parrondo in section~3.2 of \cite{HP} in the context of the multi-particle Szilard engine. 
}
\label{f:process}
\end{figure}

\section{A realization of the SHNY process}
We shall construct a concrete example of a SHNY process that consists of three steps, namely, the trapping process, the insertion of the wall, and the untrapping process, and justify the expression \rlb{Wtot} for the necessary work.
The process is different from that discussed in the original work \cite{SHNY}, and is based on the construction of Horowitz and Parrondo in section~3.2 of \cite{HP}.
See Figure~\ref{f:process}.
Since the work required in a quasi-static isothermal process depends only on the initial and the final states, it is enough to consider one example.

\subsection{The trapping Hamiltonian and the trapped state}
\begin{figure}
\centerline{\epsfig{file=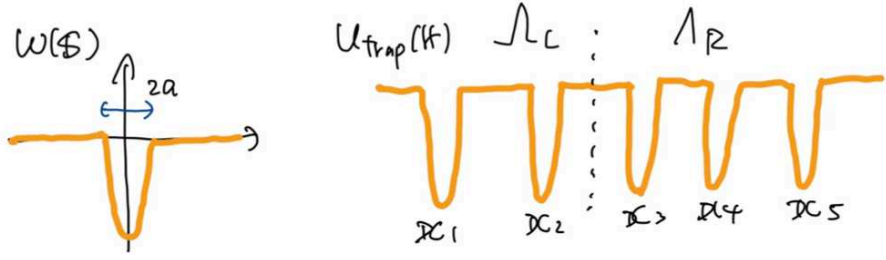,width=12truecm}}
\caption[dummy]{
The potential $w(\bss)$ for a single well and the trapping potential $\utrap(\bsr)$ which consists of five potential wells.
There is also a repulsive interaction between particles that inhibits two particles to occupy a single potential well. 
}
\label{f:trap}
\end{figure}

Let $w(\bss)\le0$ be a potential that describes a single potential well.
It has a deep unique minimum at $\bss=(0,0,0)$ and satisfies $w(\bss)=0$ for $|\bss|\ge a$, where $a>0$ is a constant.
Let $\bsx_1,\ldots,\bsx_{\NL}\in\LaL$ and $\bsx_{\NL+1},\ldots,\bsx_N\in\LaR$ be the locations of potential wells.
Note that there are $\NL$ wells in $\LaL$ and $\NR$ wells in $\LaR$.
We assume that $|\bsx_j-\bsx_k|\ge 4a$ for any $j\ne k$, and that any $\bsx_j$ is separated by a distance larger than $a$ from the wall between $\LaL$ and $\LaR$.
We then define the trapping potential $\utrap(\bsr)$ for $\bsr\in\La$ as
\eq
\utrap(\bsr)=\begin{cases}
w(\bsr-\bsx_j)&\text{if $|\bsr-\bsx_j|\le a$ for some $j$};\\
0&\text{otherwise}.
\end{cases}
\en
See Figure~\ref{f:trap}.

For $\lambda>0$, we define the trapping Hamiltonian by
\eq
\Htrap(\bsR,\bsP)=\sum_{j=1}^N\frac{|\bsp_j|^2}{2m}+\lambda\Bigl\{\sum_{j=1}^N\utrap(\bsr_j)+\frac{1}{2}\mathop{\sum_{j,k=1}^N}_{(j\ne k)}\vrep(|\bsr_j-\bsr_k|)\Bigr\},
\lb{Htrap}
\en
where the two-particle repulsion potential is defined by $\vrep(r)=0$ for $r>2a$ and $\vrep(r)=\bar{v}$ for $r\le 2a$, where $\bar{v}$ is a positive constant.
We denote the equilibrium state corresponding to \rlb{Htrap} as
\eq
\calP_{\rm trap}^{(\lambda)}(\bsR,\bsP)=\frac{e^{-\beta \Htrap(\bsR,\bsP)}}{Z_{\rm trap}^{(\lambda)}(\beta)},
\lb{rt}
\en
where
\eq
Z_{\rm trap}^{(\lambda)}(\beta)=\int d\bsR\,d\bsP\,e^{-\beta\Htrap(\bsR,\bsP)}.
\en

Observe that the potential energy $\lambda\{\sum_{j=1}^N\utrap(\bsr_j)+\frac{1}{2}\sum_{j,k=1\,(j\ne k)}^N\vrep(\bsr_j,\bsr_k)\}$ in $\Htrap(\bsR,\bsP)$ is minimized by exactly $N!$ configurations where the center of each well is occupied by one particle.
This means that, when $\lambda$ is large enough, the partition function is well approximated by only taking into account particle configurations in which each well contains exactly one particle.
We thus get
\eq
Z_{\rm trap}^{(\lambda)}(\beta)\simeq N!\,\bigl\{z_{\rm trap}^{(\lambda)}(\beta)\bigr\}^N,
\lb{Zz}
\en
where $N!$ counts the possible assignments of particles to the wells, and
\eq
z_{\rm trap}^{(\lambda)}(\beta)=\int d^3\bsr\,d^3\bsp\,\exp\bigl[-\beta\bigl\{\tfrac{|\bsp|^2}{2m}+\lambda\, w(\bsr)\bigr\}\bigr]
\en
is the single-particle partition function of a system consists of a single potential well.
Correspondingly the equilibrium state $\calP_{\rm trap}^{(\lambda)}(\bsR,\bsP)$ is approximated as
\eq
\calP_{\rm trap}^{(\lambda)}(\bsR,\bsP)\simeq\frac{1}{N!}\sum_{\uppi}
\prod_{j=1}^N\frac{\exp[-\beta\{\frac{|\bsp_{\uppi(j)}|^2}{2m}+\lambda\, w(\bsr_{\uppi(j)}-\boldsymbol{x}_j)\}]}{z_{\rm trap}^{(\lambda)}(\beta)},
\lb{rt2}
\en
where $\uppi$ is summed over $N!$ permutations of particle labels $j=1,\ldots,N$.

Note that possible errors in the approximate formulas \rlb{Zz} and \rlb{rt2} are of order $e^{-A\lambda}$, where $A$ is a constant.
Since the errors can be made practically zero by making $\lambda$ large enough, we shall treat  \rlb{Zz} and \rlb{rt2} (and similar expressions) as equalities in the rest of the note.

\subsection{The trapping process}
\label{s:trap}
Let $\alpha\in[0,1]$ and suppose that our system is described by the interpolating Hamiltonian
\eq
H_\alpha(\bsR,\bsP)=(1-\alpha)\,H_0(\bsR,\bsP)+\alpha\,\Htrap(\bsR,\bsP).
\lb{Ha}
\en
We first set $\alpha=0$ and assume that the system is in the corresponding equilibrium state \rlb{r0}.
We place the system in contact with a heat bath at inverse temperature $\beta$ and slowly change $\alpha$ from 0 to 1.
Then a quasi-static isothermal operation is realized, and we end up in the equilibrium state \rlb{rt}.

From the basic formula \rlb{WZZ} we see that the work done to the system during this process is written in terms of the initial and the final partition functions as 
\eq
W_{\rm trap}=\frac{1}{\beta}\log Z_0(\beta)-\frac{1}{\beta}\log Z_{\rm trap}^{(\lambda)}(\beta).
\lb{Wt0}
\en
By using \rlb{Zz}, which we now regard as equality, we obtain
\eq
W_{\rm trap}=\frac{1}{\beta}\log Z_0(\beta)-\frac{1}{\beta}\log N!\,\bigl\{z_{\rm trap}^{(\lambda)}(\beta)\bigr\}^N
=\frac{1}{\beta}\log\frac{Z_0(\beta)}{N!} -\frac{N}{\beta}\log z_{\rm trap}^{(\lambda)}(\beta).
\lb{Wt}
\en

\subsection{The insertion of the wall and the relabeling}
\label{s:wall}
In the next step, we insert the infinitesimally thin planar wall into $\La$ and divide it into $\LaL$ and $\LaR$.
Since we only need to consider particle configurations in which every particle is trapped in one of the trapping wells, the insertion of the wall does not change the state \rlb{rt} or, equivalently, \rlb{rt2} at all.
We thus see that the work associated with the wall insertion can be neglected.
Moreover, we find that this process is quasi-static\footnote{%
We here require that a quasi-static process should be reversible in the sense that one can always undo the process.
In this case, the reversibility is obvious since one can freely insert or remove the wall without changing the state.
We nevertheless note that some experts may be reluctant to call the process quasi-static since it does not belong to the standard class of processes realized by the slow modification of the Hamiltonian.

We note that the corresponding process proposed by Sasa, Hiura, Nakagawa, and Yoshida, which they call the quasi-static decomposition, only makes use of a continuous change of soft (but long-ranged) interactions.
} as well as isothermal.

\begin{figure}
\centerline{\epsfig{file=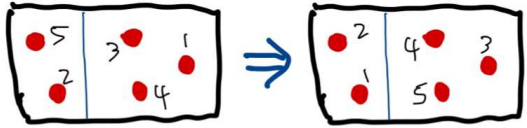,width=7truecm}}
\caption[dummy]{
An example of the redefinition of the particle labels.
Note that multiple configurations are mapped into a single configuration in this procedure.
Nevertheless, the procedure can be reversed by virtue of the identity of the particles.
}
\label{f:relabel}
\end{figure}

At this stage, it is theoretically convenient to redefine the labels of particles.
Given a particle configuration, let $j_1,\ldots,j_{\NL}$ be the labels of particles in $\LaL$ with the ordering $j_{q}<j_{q+1}$.
We then redefine labels so that $j_1,\ldots,j_{\NL}$ are replaced (with this order) by $1,\ldots,\NL$.
Similarly let $k_1,\ldots,k_{\NR}$ be the labels of particles in $\LaR$ with $k_1<k_{q+1}$.
We then replace $k_1,\ldots,k_{\NR}$ with $\NL+1,\ldots,N$.
See Figure~\ref{f:relabel}.

Note that exactly $N!/(\NL!\NR!)$ apparently different particle configurations are mapped into a single configurations in this relabeling.
This is of course not a problem since all these configurations are physically identical.
It is also easily found by inspection that the process can be reversed.
In the reversed process one generates $N!/(\NL!\NR!)$ different particle configurations with equal weights from a single configuration.

After the relabeling, the state \rlb{rt2} becomes a product state
\eq
\calP_{\rm trap,LR}^{(\lambda)}(\bsR,\bsP)=\calP_{\rm trap,L}^{(\lambda)}(\bsR_{\rm L},\bsP_{\rm L})\,\calP_{\rm trap,R}^{(\lambda)}(\bsR_{\rm R},\bsP_{\rm R}),
\en
with
\eqg
\calP_{\rm trap,L}^{(\lambda)}(\bsR_{\rm L},\bsP_{\rm L})
=\frac{1}{\NL!}\sum_{\uppi_{\rm L}}
\prod_{j=1}^{\NL}\frac{\exp[-\beta\{\frac{|\bsp_{\uppi_{\rm L}(j)}|^2}{2m}+\lambda\, w(\bsr_{\uppi_{\rm L}(j)}-\boldsymbol{x}_j)\}]}{z_{\rm trap}^{(\lambda)}(\beta)},\\
\calP_{\rm trap,R}^{(\lambda)}(\bsR_{\rm R},\bsP_{\rm R})
=\frac{1}{\NR!}\sum_{\uppi_{\rm R}}
\prod_{j=\NL+1}^{N}\frac{\exp[-\beta\{\frac{|\bsp_{\uppi_{\rm R}(j)}|^2}{2m}+\lambda\, w(\bsr_{\uppi_{\rm R}(j)}-\boldsymbol{x}_j)\}]}{z_{\rm trap}^{(\lambda)}(\beta)},
\eng
where $\uppi_{\rm L}$ and $\uppi_{\rm R}$ are summed over all permutations of $\{1,\ldots,\NL\}$ and $\{\NL+1,\ldots,N\}$, respectively.
Recalling the relation between the expressions \rlb{rt} and \rlb{rt2} of the trapped equilibrium states in the whole lattice, one immediately sees that the above states are also written as
\eq
\calP_{\rm trap,L}^{(\lambda)}(\bsR_{\rm L},\bsP_{\rm L})=\frac{H_{\rm trap,L}^{(\lambda)}(\bsR_{\rm L},\bsP_{\rm L})}{Z_{\rm trap,L}^{(\lambda)}(\beta)},
\quad
\calP_{\rm trap,R}^{(\lambda)}(\bsR_{\rm R},\bsP_{\rm R})=\frac{H_{\rm trap,R}^{(\lambda)}(\bsR_{\rm R},\bsP_{\rm R})}{Z_{\rm trap,R}^{(\lambda)}(\beta)},
\lb{rt5}
\en
where
\eqg
H_{\rm trap,L}^{(\lambda)}(\bsR_{\rm L},\bsP_{\rm L})=\sum_{j=1}^{\NL}\frac{|\bsp_j|^2}{2m}+\lambda\Bigl\{\sum_{j=1}^{\NL}\utrap(\bsr_j)+\frac{1}{2}\mathop{\sum_{j,k=1}^{\NL}}_{(j\ne k)}\vrep(|\bsr_j-\bsr_k|)\Bigr\},
\\
H_{\rm trap,R}^{(\lambda)}(\bsR_{\rm R},\bsP_{\rm R})=\sum_{j=\NL+1}^{N}\frac{|\bsp_j|^2}{2m}+\lambda\Bigl\{\sum_{j=\NL+1}^N\utrap(\bsr_j)+\frac{1}{2}\mathop{\sum_{j,k=\NL+1}^{N}}_{(j\ne k)}\vrep(|\bsr_j-\bsr_k|)\Bigr\},
\eng
are nothing but the trapping Hamiltonian \rlb{Htrap} restricted onto $\LaL$ and $\LaR$, respectively.

\subsection{The untrapping process}
Finally, we consider the untrapping process, which is the exact opposite of the trapping process discussed in section~\ref{s:trap}, separately for the systems in $\LaL$ and $\LaR$.
To be precise we define for $\alpha\in[1,2]$ the new interpolating Hamiltonians
\eqg
H_{\alpha, \rm L}(\bsR_{\rm L},\bsP_{\rm L})=(2-\alpha)\,H_{\rm trap,L}^{(\lambda)}(\bsR_{\rm L},\bsP_{\rm L})+(\alpha-1)\,H_{\rm L}(\bsR_{\rm L},\bsP_{\rm L}),\\
H_{\alpha, \rm R}(\bsR_{\rm R},\bsP_{\rm R})=(2-\alpha)\,H_{\rm trap,R}^{(\lambda)}(\bsR_{\rm R},\bsP_{\rm R})+(\alpha-1)\,H_{\rm R}(\bsR_{\rm R},\bsP_{\rm R}),
\eng
and change $\alpha$ slowly from 1 to 2.
Since the two systems start from the trapped states \rlb{rt5}, they end up in the desired equilibrium states \rlb{rLrR}.

The works done to the systems in these quasi-static isothermal processes are the opposite of the work for the trapping processes.
Since \rlb{Wt} gives a general expression for the trapping work, we readily find
\eqg
W_{\rm untrap,L}=-\frac{1}{\beta}\log\frac{Z_{\rm L}(\beta)}{\NL!} +\frac{\NL}{\beta}\log z_{\rm trap}^{(\lambda)}(\beta),\lb{WL}\\
W_{\rm untrap,R}=
-\frac{1}{\beta}\log\frac{Z_{\rm R}(\beta)}{\NR!}+\frac{\NR}{\beta}\log z_{\rm trap}^{(\lambda)}(\beta).
\lb{WR}
\eng

\subsection{The total work in the SHNY process}
Thus, by combining the trapping process, the insertion of the wall, and the untrapping processes, we constructed a SHNY process, i.e., a quasi-static isothermal process that brings the state of the whole system from the initial equilibrium state \rlb{r0} to the final separated equilibrium state \rlb{rLR}.
From the expressions \rlb{Wt}, \rlb{WL}, and \rlb{WR} for the trapping and untrapping works, we immediately find
\eqa
W_{\rm SHNY}&=W_{\rm trap}+W_{\rm untrap,L}+W_{\rm untrap,R}
\nl&=\frac{1}{\beta}\log\frac{Z_0(\beta)}{N!}-\Bigl\{\frac{1}{\beta}\log\frac{Z_{\rm L}(\beta)}{\NL!}
+\frac{1}{\beta}\log\frac{Z_{\rm R}(\beta)}{\NR!}\Bigr\},
\lb{W}
\ena
which is the desired expression \rlb{Wtot} for the whole isothermal work.

\section{Alternative formulation}
Let us present a different derivation of the main result \rlb{W} on the total work for the SHNY process.
This derivation is indeed closer in spirit to that in the original work \cite{SHNY}.
Here we do not change particle labels throughout the isothermal processes.

We again start from the equilibrium state \rlb{r0}, perform the trapping process, and then insert the wall.
We then rewrite the state \rlb{rt} in order to emphasize the presence of the wall as
\eq
\widetilde{\calP}_{\rm trap}^{(\lambda)}(\bsR,\bsP)=
\begin{cases}
\dfrac{e^{-\beta \Htrap(\bsR,\bsP)}}{Z_{\rm trap}^{(\lambda)}(\beta)}&\text{if $\bsR\in\CLR$};\\
0&\text{otherwise}.
\end{cases}
\lb{trr}
\en
Here $\CLR$ denotes the subset of $\La^N$ in which there are $\NL$ particles in the region $\LaL$ and $\NR$ particles in $\LaR$.

With the same $u_1(\bsr)$ and $v_1(\bsr,\bsr')$ as above, we define the final Hamiltonian on the whole region $\La$ as
\eq
H_{\rm LR}(\bsR,\bsP)=\sum_{j=1}^N\frac{|\bsp_j|^2}{2m}+\sum_{j=1}^Nu_1(\bsr_j)+\frac{1}{2}\mathop{\sum_{j,k=1}^N}_{(j\ne k)}v_1(\bsr_j,\bsr_k).
\lb{HLR}
\en
We then define for $\alpha\in[1,2]$ the interpolating Hamiltonian
\eq
H_\alpha(\bsR,\bsP)=(2-\alpha)\,\Htrap(\bsR,\bsP)+(\alpha-1)\,H_{\rm LR}(\bsR,\bsP),
\en
and change $\alpha$ slowly from 1 to 2, keeping the system in touch with a heat bath at $\beta$.
Since we start from \rlb{trr} at $\alpha=1$, we end up with the state
\eq
\widetilde{\calP}_{\rm LR}(\bsR,\bsP)=
\begin{cases}
\dfrac{e^{-\beta H_{\rm LR}(\bsR,\bsP)}}{\ZLR(\beta)}&\text{if $\bsR\in\CLR$};\\
0&\text{otherwise},
\end{cases}
\lb{trLR}
\en
which is the same as the state \rlb{rLR} except for particle labels.
The partition function in \rlb{trLR} is given by
\eq
\ZLR(\beta)=\int_{\bsR\in\CLR}d\bsR\,d\bsP\,e^{-\beta H_{\rm LR}(\bsR,\bsP)}=\frac{N!}{\NL!\NR!}\,Z_{\rm L}(\beta)\,Z_{\rm R}(\beta),
\lb{ZZZ}
\en
where the final identity follows by examining the assignments of particle labels.

Since the whole process is quasi-static and isothermal, we find from the general formula \rlb{WZZ} that the work done to the system is
\eq
W_{\rm SHNY}=\frac{1}{\beta}\log Z_0(\beta)-\frac{1}{\beta}\log\ZLR(\beta)
=\frac{1}{\beta}\log\frac{Z_0(\beta)}{N!}-\Bigl\{\frac{1}{\beta}\log\frac{Z_{\rm L}(\beta)}{\NL!}
+\frac{1}{\beta}\log\frac{Z_{\rm R}(\beta)}{\NR!}\Bigr\},
\lb{W2}
\en
which of course coincides with \rlb{Wtot} and \rlb{W}.

\section{The role of the identity of particles}
The fully operational point of view of Sasa, Hiura, Nakagawa, and Yoshida presented above allows us to answer various ``frequently asked questions'' in a definitive and clear manner.
Let us discuss two typical questions.

\paragraph{The meaning of the identity of particles}
A natural question is what does it exactly mean that classical particles are identical to each other.
It is clear from the foregoing discussion that all that we need are particles that have common mass $m$, common single-particle potential, and common interaction potential.
In this sense, the identity is of a purely dynamical nature.
In particular, it is conceptually different from that of quantum systems, which is directly related to the basic structure of the underlying many-particle Hilbert space.

In our discussion, the identity in this classical sense is used explicitly in section~\ref{s:wall}, where we redefined particle labels.
It is crucial to note that one may redefine particle labels of non-identical particles, but the relabeling process then becomes irreversible.
To see this, imagine an extreme case where there is a potential or interaction that distinguishes between the particles, and it happens with a large probability that the particles $1,\ldots,\NL$ are found in $\LaL$ and the particles $\NL+1,\ldots,N$ are found in $\LaR$ immediately before (and hence after) the wall insertion.
In this case, no changes are made in the relabeling process.
But if we consider the reversed process of this relabelling, the only thing we can do is to generate $N!/(\NL!\NR!)$ different labelings with equal probabilities.
We then end up in a different state than the original.

\paragraph{Identical but distinguishable particles}
Another question deals with a situation where the particles are identical in the above dynamical sense but are indeed distinguishable.

Suppose, for example, that there are $\NL$ red particles and $\NR$ green particles, but all of them are governed by the Hamiltonian \rlb{H}.
One can then take the separated equilibrium state \rlb{rLR} where all particles in $\LaL$ are red and all particles in $\LaR$ are green.
We now perform the inverse of our SHNY process and bring the state to the equilibrium state \rlb{r0}, in which green and red particles are mixed in a completely random manner.
If we further perform our SHNY process (in the forward direction), then the system will go back to the same state \rlb{rLR} but with mixed colors (unless $N$ is too small).
In this sense, the first process is clearly irreversible.
Can we regard it as a quasi-static isothermal process?

The answer is yes, provided that all potentials and interactions that we can make use of are independent of the particle colors.
We can simply ignore the colors of particles and can repeat the operational consideration of Sasa, Hiura, Nakagawa, and Yoshida to get the same conclusion \rlb{PNfinal} from the basic premise \rlb{SHNY}.

But the very fact that we can see the colors of particles suggests that there are some potentials or interactions that behave differently for green and red particles.
If such potentials or interactions are available, then we can construct processes in which green and red particles are not identical.
The foregoing discussion leading to \rlb{PNfinal} is no longer meaningful.
As this simple example illustrates, the identity of particles in the operational framework may depend on the resources that we, the experimenters, have.

\bigskip
{\small
It is a pleasure to thank Naoko Nakagawa and Shin-ichi Sasa for their indispensable discussions.
The present work was supported by JSPS Grants-in-Aid for Scientific Research No. 22K03474.
}

\end{document}